# GEANT4 SIMULATION OF THE BREMSSTRAHLUNG CONVERTER OPTIMAL THICKNESS FOR STUDYING THE RADIATION DAMAGE PROCESSES IN ORGANIC DYES SOLUTIONS

Tetiana V. Malykhina[a,b]*, Vladimir E. Kovtun[a], Valentin I. Kasilov[b], Sergey P. Gokov[b]

[a]*Kharkiv V.N. Karazin National University*
*4, Svobody sq., 61022, Kharkiv, Ukraine*
[b]*National Science Center "Kharkiv Institute of Physics and Technology"*
*2, Academicheskaya str., 61108, Kharkiv, Ukraine*
*Corresponding Author: malykhina@karazin.ua

The study of the processes occurring in a matter when ionizing radiation passes through is important for solving various problems. Examples of such problems are applied and fundamental problems in the field of radiation physics, chemistry, biology, medicine and dosimetry. This work is dedicated to computer modeling of the parameters of a tungsten converter for studying the processes of radiation damage during the interaction of ionizing radiation with solutions of organic dyes. Simulation was carried out in order to determine the optimal thickness of the converter under predetermined experimental conditions. Experimental conditions include: energies and type of primary particles, radiation intensity, target dimensions, relative position of the radiation source and target. Experimental studies of the processes of radiation damage occurring in solutions of organic dyes are planned to be carried out using the linear electron accelerator "LINAC-300" of the National Scientific Center "Kharkov Institute of Physics and Technology". Electrons with 15 MeV energy are chosen as primary particles. The interaction of electrons with the irradiated target substances is planned to be studied in the first series of experiments. Investigations of the interaction of gamma quanta with the target matter will be carried out in the second series of experiments. The tungsten converter is used to generate a flux of bremsstrahlung gamma rays. One modeling problem is determination of the converter thickness at which the flux of bremsstrahlung gamma will be maximal in front of the target. At the same time, the flow of electrons and positrons in front of the target should be as low as possible. Another important task of the work is to identify the possibility of determining the relative amount of radiation damage in the target material by the Geant4-modeling method. Radiation damage of the target substance can occur due to the effect of bremsstrahlung, as well as electrons and positrons. Computational experiments were carried out for various values of the converter thickness – from 0 mm (no converter) to 8 mm with a step of 1 mm. A detailed analysis of the obtained data has been performed. As a result of the data analysis, the optimal value of the tungsten converter thickness was obtained. The bremsstrahlung flux in front of the target is maximum at a converter thickness of 2 mm. But at the same time, the flux of electrons and positrons crossing the boundaries of the target does not significantly affect the target. The computational experiment was carried out by the Monte Carlo method. A computer program in C++ that uses the Geant4 toolkit was developed to perform calculations. The developed program operates in a multithreaded mode. The multithreaded mode is necessary to reduce the computation time when using a large number of primary electrons. The G4EmStandardPhysics_option3 model of the PhysicsList was used in the calculations. The calculations necessary for solving the problem were carried out using the educational computing cluster of the Department of Physics and Technology of V.N. Karazin Kharkiv National University.

Linear electron accelerators are currently used to solve various applied and fundamental problems. In particular, accelerators are used in nuclear medicine, materials science, in the development of radiation detectors, etc. Experiments on the irradiation of organic dyes were carried out in [1]. Aqueous, alcoholic and glycerol solutions of methylene blue ($C_{16}H_{18}N_3SCl$), as well as methyl orange ($C_{14}H_{14}N_3O_3SNa$), were irradiated with an electron beam. The authors investigated the optical density of the irradiated dyes. The electron beam energy was 16 MeV. It was found that aqueous solutions have less radiation resistance than alcohol and glycerol solutions. Further experimental studies of radiation-stimulated chemical processes during the destruction of dyes organic molecules have become necessary.

An experimental stand with a bremsstrahlung converter is usually used to study the effects that occur in matter when interacting with a flux of gamma quanta. The generation of a gamma quanta flux occurs in the converter due to the conversion of part of the electron beam energy into bremsstrahlung. The converter must be made of a material with a high atomic number and a high density in order to obtain a sufficient amount of gamma quanta. Tantalum and tungsten are appropriate materials due to their physical and chemical properties [2].

Computer simulation using Geant4 allows virtual nuclear physics experiments. These experiments are necessary for preliminary assessment, as well as choice of the bremsstrahlung converter optimal parameters for real experiments.

The aim of this work is to select the optimal thickness of the bremsstrahlung converter as part of the experimental stand "LINAC-300" of the National Scientific Center "Kharkov Institute of Physics and Technology" for studying the radiation resistance of organic dye solutions. The primary electron energy is 15 MeV. It is necessary to determine the optimal thickness of the tungsten converter, at which the bremsstrahlung flux will be maximum directly in front of the target containing the solution. At the same time, the flux of electrons and positrons should be as low as possible. A converter with these parameters will be necessary to study the nature of the mechanisms that lead to radiation damage that occurs when ionizing radiation interacts with organic dye solutions.

**Materials and methods**

The scheme of the planned experiment is shown in Figure 1. The primary electrons beam passes through the titanium outlet window foil of the LINAC accelerator. The electron energy is $E_e=15$ MeV. The thickness of the foil is 50 μm. The bremsstrahlung converter is located at a distance of 50 mm from the titanium foil. The converter is marked in black in Figure 1. The electrons beam is directed along the normal to the converter surface. The target containing the solution is located after the converter. The target is marked in light green in Figure 1. The irradiated target contains 1% aqueous solution of an organic dye. Methylene blue $C_{16}H_{18}N_3SCl$ was chosen as an organic dye, as one of the substances studied in [1].

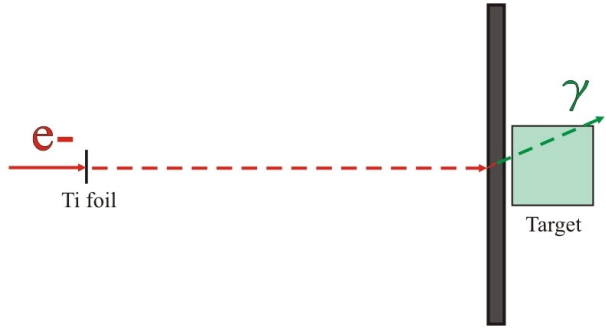

**Figure 1.** Simplified scheme of the experiment. The target size is 10 mm × 10 mm × 10 mm

The dimensions of the target are 10 mm×10 mm×10 mm, i.e., rather small, due to the need for further studies of the target. The transverse dimensions of the converter are 40 mm × 40 mm. The irradiated target is located at a distance of 1 mm after the converter. The thickness of the tungsten converter is varied from 0 mm to 8 mm in 1 mm increment in a series of computational experiments. A value of 0 mm corresponds to the case when the converter is absent. This case is necessary to study the interaction of electrons with the target material in the absence of bremsstrahlung. In this case, there will still be a small amount of gamma quanta formed in the titanium foil material when primary electrons pass through it.

One of the objectives of this work is to determine the thickness of the converter at which the flux of bremsstrahlung gamma quanta immediately in front of the target will be maximum, regardless of the amount of electrons and positrons in front of the target. The number of electrons and positrons in front of the target in this case should be minimal. It is necessary to estimate the relative amount of radiation damage in the target containing the organic dye solution for each value of the tungsten converter thickness.

We have developed a computer model of the planned experiment to solve the problem. The model is based on a computer program. The program is developed in C++ and uses the Geant4 toolkit of version 10.6 [3, 4].

The Geant4 toolkit has a complete set of tools for computer modeling of nuclear physical processes of radiation with matter interaction. Geant4 modules used in our program include Instrumentation for describing the detectors geometry and the experimental setup as a whole, description of particles and physical processes, transport and tracking of particles, simulation of the detector response. The Geant4 library uses CLHEP classes [5] and has a wide range of utility functions as well as random number generators.

The developed program contains the definition of several main classes that correspond to the specifics of the task when using the Geant4 library. All these classes must be registered in a special object-instance of the G4RunManager class, which controls the modeling process. The main classes are G4VUserDetectorConstruction, G4VPhysicsList, G4VUserPrimaryGeneratorAction [3]. The G4VUserDetectorConstruction class contains the geometry definition of the experimental setup model and its constituent parts, their mutual arrangement, as well as their materials. The G4VPhysicsList class is necessary to describe the models of physical processes [4] that occur during the interaction of ionizing radiation with materials of experimental setup components. We specify the primary particles source in the simulation, their type and energies in the G4VUserPrimaryGeneratorAction class. We also specify the particles movement direction and other parameters that characterize the radiation source in the G4VUserPrimaryGeneratorAction class. In addition, optional classes, for example, G4UserEventAction, G4UserSteppingAction [3], etc. can also be registered in the instance of the G4RunManager class. These classes allow one to control the behavior of the developed program at various stages of its execution, as well as set the required level of detail when displaying results.

The Geant4 toolkit developers offer about 30 predefined PhysicsList models to date. These models are described in detail in the documentation [6], and have application in modeling of almost any problem - from problems in high energy physics to applications in microdosimetry. We have chosen the G4EmStandardPhysics_option3 model of the PhysicsList module to solve our problem. This model is the most suitable [6] for simulating the passage of 15 MeV electron beam through the blocks of the studied facility. A monoenergetic electron beam is used in our problem to simulate ideal experimental conditions. The electron beam diameter in the model is specified to be 2 mm due to the fact that this series of calculations is preliminary. We will be able to find out the real parameters of the beam after carrying out a real experiment in order to use them in further calculations. The particle transport-tracking threshold was specified to be 0.1 mm in length units. We used the setCut() function to calculate the threshold for tracking particle transport in units of energy. The threshold is approximately 351 keV for electrons in tungsten, and 36 keV for gamma rays in tungsten. The tracking threshold for particle transport is, respectively, 85 keV for electrons and 1 keV for gamma quanta in an aqueous solution of methylene blue.

The program we have developed contains a visualization module that uses the OpenGL library. A screen shot of the visualization module is shown in Figure 2 (a, b) in order to visualize the relative position of the experimental setup components, as well as to demonstrate the result of the particles passage through the converter.

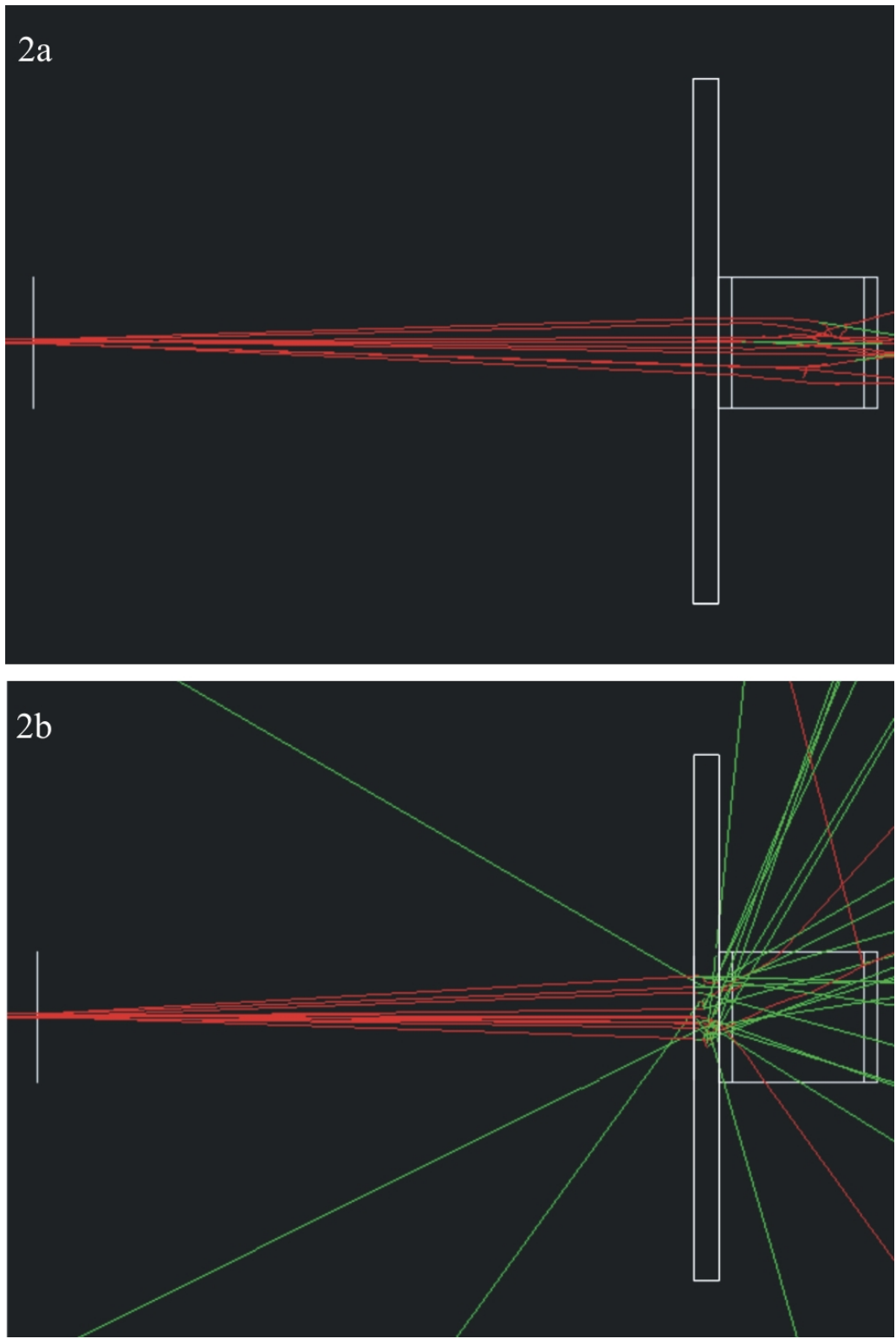

**Figure 2.** Passage of 10 primary electrons through the facility. The electron energy is 15 MeV.
(Figure 2a – The converter is absent; Figure 2b – We use a bremsstrahlung converter, its thickness is 2 mm)

Figure 2 shows the converter with a thickness of 2 mm. The converter thickness increases by 1 mm in each subsequent series of computational experiments. The energy of primary electrons is 15 MeV. The trajectories of electrons are shown in red in Figure 2, the trajectories of bremsstrahlung gamma quanta are shown in green. The

OpenGL visualization of 10 primary electrons passing through the setup is shown in Figure 2a. There is no bremsstrahlung converter in this case, but the outline of the converter is shown in the screen shot in order to to keep the scale. We use air instead of a tungsten converter in the DetectorConstruction module in this case. A slight deviation of the electron beam can be seen after passing through the thin titanium foil. All primary electrons hit the target.

The passage of 10 electrons through a tungsten bremsstrahlung converter is shown in Figure 2b. The trajectories of gamma quanta are shown in green, the trajectories of electrons are shown in red in Figure 2b, similar to the marking in Figure 2a. It can be seen that primary electrons are decelerated in the converter, and bremsstrahlung gamma quanta are formed. A certain amount of gamma quanta does not hit the target in this case due to the small size of the target. Therefore, calculations using a large number of events are necessary, as well as an analysis of the obtained results.

The program developed by us has a batch mode for performing calculations by the Monte Carlo method for the purpose of further statistical data processing. The passage of $10^7$ primary electrons with energy $E_e$=15 MeV through the setup containing the target was simulated in the batch mode. The statistical error of the Monte Carlo method will be less than 1% for this amount of primary electrons.

The calculations required to solve the problem were carried out using the educational computing cluster of the Scientific and Educational Institute "Physics and Technology Faculty" of V.N. Karazin Kharkiv National University. The educational compute cluster (Figure 3) consists of Dell Power Edge 1850 blocks of various configurations. These blocks are integrated into a local network, and use the Linux operating system.

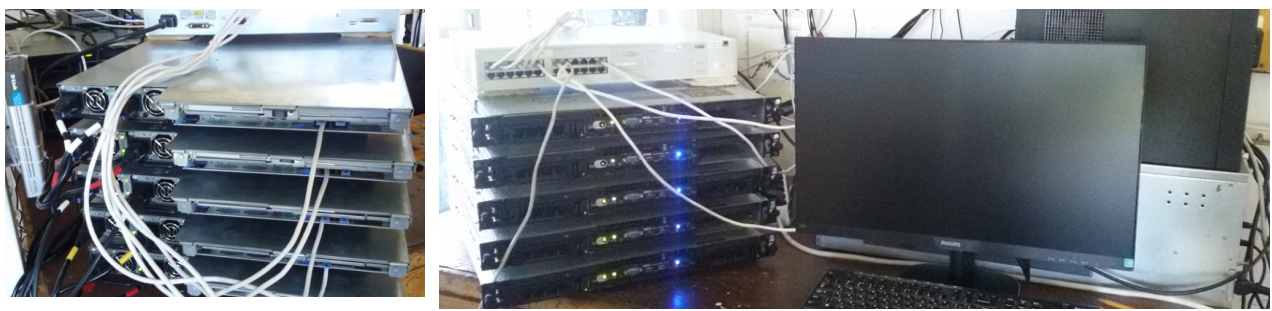

**Figure 3**. Educational computing cluster of the Scientific and Educational Institute "Physics and Technology Faculty"

We used the multi-threaded mode of the Geant4 toolkit. The required additional libraries have been installed for this. In addition, we modified the source codes of the G4MPI module [3, 7] in order to uniformly load the computing blocks of the cluster, depending on their performance.

**Results and discussion**

The series of computational experiments were carried out using the batch mode of the developed program. The energy spectra of bremsstrahlung gamma quanta (Figure 4) in front of the target containing an organic dye solution were obtained as a result of data processing. The values of the gamma quanta flux directly in front of the target were also calculated for different converter thicknesses (Figure 5). The values of the electron flux in front of the target (Figure 6) were obtained depending on the thickness of the converter. In addition, the energy spectra of electrons in front of the target were calculated for two values of the converter thickness (Figure 7), at which a significant flux of gamma quanta is observed in front of the target. All presented results are normalized to 1 incident electron. The transverse dimension of the target is 10 mm×10 mm. The transverse dimension of the converter is 40 mm×40 mm.

The energy spectra of bremsstrahlung gamma quanta for primary electrons with energy $E_e$ = 15 MeV in front of the target are shown in Figure 4. These spectra have 100 keV resolution and are presented in a log scale on Y-axis. It can be noted that the maximum relative amount of gamma rays crossing the target boundary will occur for tungsten converter thickness of 2 mm.

The relative amount of bremsstrahlung gamma quanta crossing the target boundary, depending on the converter thickness, is shown in Figure 5. The flux is normalized to 1 incident electron. It can be seen that the maximum flux of bremsstrahlung gamma quanta in front of the target will be at the converter thickness of 2 mm for primary electrons with energy of 15 MeV. The flux of gamma quanta decreases when the converter thickness is 3 mm.

The electron flux values in front of the target (Figure 6) were obtained for the same values of the converter thickness from 0 mm to 8 mm with a step of 1 mm.

It can be seen that with a converter thickness of 2 mm, when the largest amount of bremsstrahlung gamma quanta is observed in front of the target, there is also a significant amount of electrons in front of the target. The relative electrons number in front of the target, depending on the converter thickness, is presented in Table 1. The presented results are normalized to 1 incident electron.

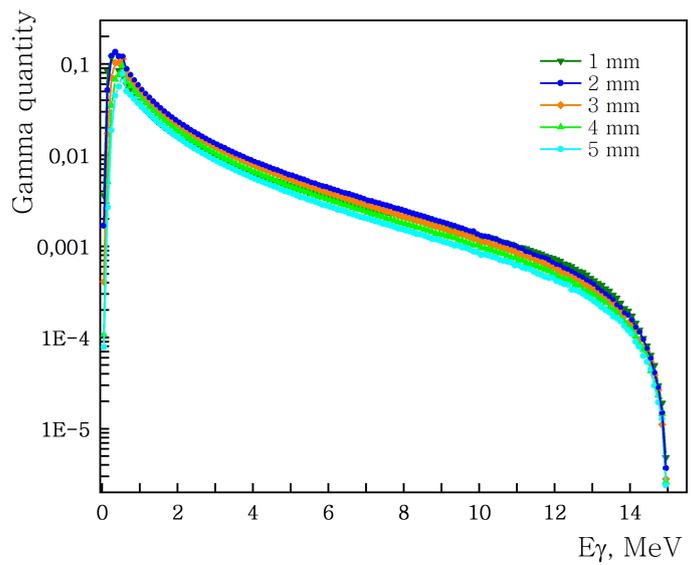
**Figure 4.** Energy spectra of gamma quanta crossing the target boundary

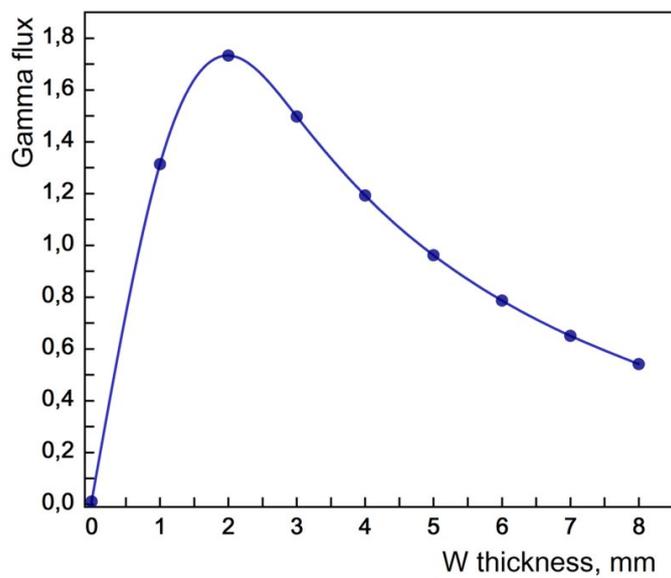
**Figure 5.** The gamma quanta flux crossing the target boundaries, depending on the converter thickness

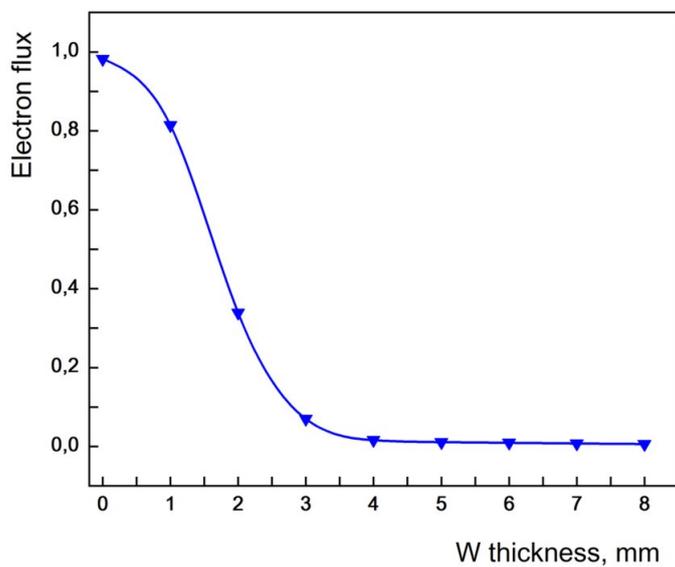
**Figure 6.** The electrons flux crossing the target boundaries, depending on the converter thickness

**Table 1.**
Relative number of electrons crossing the target boundary, depending on the bremsstrahlung converter thickness for primary electrons with energy $E_e$ = 15 MeV

| Tungsten thickness, mm | 0 | 1 | 2 | 3 | 4 | 5 | 6 | 7 | 8 |
|---|---|---|---|---|---|---|---|---|---|
| Amount of electrons in front of the target | 0.982 | 0.815 | 0.339 | 0.0703 | 0.0162 | 0.0112 | 0.0092 | 0.0076 | 0.0064 |

Comparison of the electrons energy spectra in front of the target was carried out for two values of the tungsten converter thickness. These studies are necessary to choose the optimal converter thickness, at which it is expedient to study the mechanisms of radiation damage in the target. Two values of thickness for the tungsten converter are 2 mm and 3 mm (Figure 7). The calculations of spectra were done in same steps of 1 MeV for the convenience of results comparing.

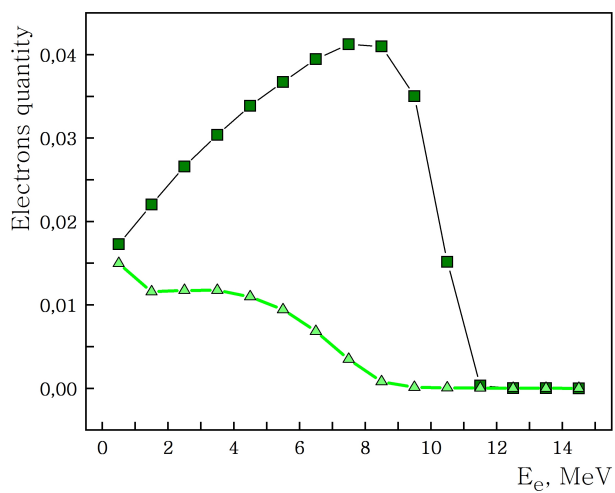

**Figure 7.** Comparison of the electrons energy spectra in front of the target for two values of the tungsten converter thickness. These values are 2 mm and 3 mm. The energy of primary electrons is $E_e$ = 15 MeV. The electrons spectrum in the case of 2 mm converter is indicated by dark squares. The electrons spectrum in the case of 3 mm converter is indicated by light triangles

It can be seen (Figure 7) that a significant amount of high-energy electrons are still present in front of the target with a converter thickness of 2 mm. However, the energies of electrons in front of the target noticeably decrease with increasing the converter thickness to 3 mm. The relative amount of electrons also decreases. The electrons energy spectra (Figure 7) are normalized to 1 incident electron. A significant decreasing the number of electrons in front of the target, as well as decreasing the electron energy, suggests us that it is desirable to use a 3 mm thick tungsten bremsstrahlung converter irradiated with 15 MeV primary electrons to study the radiation damages processes in organic solutions irradiated by gamma quanta.

The simulation of the $^{16}O$ nuclei formation in the target was carried out in order to determine the possibility of a preliminary assessment of radiation damage occurring in a 1% aqueous solution of an organic dye upon irradiation by primary electrons. We took into account $^{16}O$ nuclei with energies above 5 eV. This simulation was carried out for various values of the tungsten converter thickness. The dependence of the number of $^{16}O$ nuclei with energies above 5 eV on the converter thickness is shown in Figure 8.

It can be noted that the rupture of water molecules' chemical bonds is practically not observed in the absence of bremsstrahlung gamma quanta, i.e., with a converter thickness of 0 mm. The relative number of such events is approximately $3.7 \times 10^{-5}$. A significant increase in radiation-stimulated chemical processes of the target substance destruction [8] is observed with increasing the number of gamma quanta (Figure 4 and Figure 5) crossing the target boundaries. The number of water molecules chemical bonds rupture is $5.4 \times 10^{-4}$ per one primary electron with energy of 15 MeV at 1 mm tungsten thickness. The relative number of ruptures of water molecules chemical bonds is now about 10 times greater than in case of absent converter. Approximately $6 \times 10^{-4}$ ruptures of water molecules chemical bonds per 1 incident electron occur when tungsten is 2 mm thick.

The attenuation of the gamma-ray beam occurs due a further increase in the converter thickness [9]. This fact explains decreasing (Figure 8) of the number of water molecules chemical bonds ruptures while increasing the converter thickness.

Therefore, further experimental studies of the interaction of radiation with organic dye solutions are necessary to reveal the mechanisms leading to radiation damage of dye solutions.

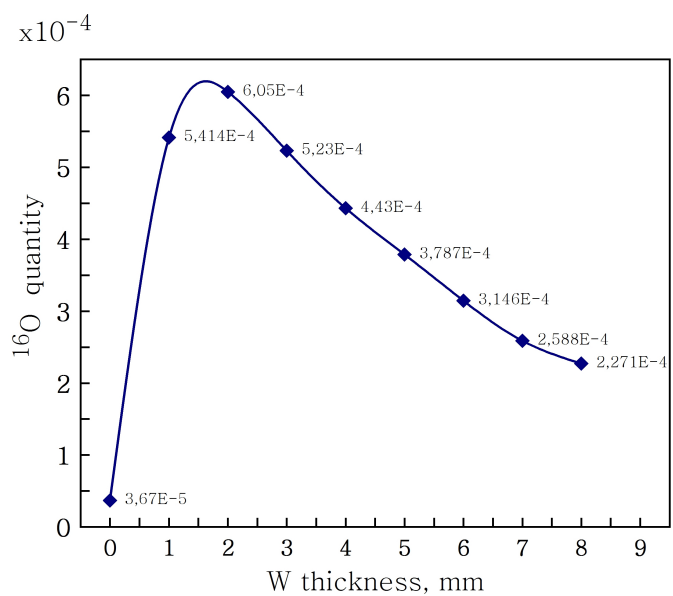

**Figure 8**. Formation of $^{16}$O nuclei with energies above 5 eV depending on the thickness of the converter

## CONCLUSIONS

The simulation of the passage of electrons flux with energy of 15 MeV through tungsten converters of various thickness values, from 0 mm to 8 mm, was carried out with a step of 1 mm. The values of the bremsstrahlung gamma flux immediately in front of a small target containing 1% aqueous solution of an organic dye were obtained as a result of simulation. It is shown that it is necessary to locate the target at the minimum possible distance from the converter due to the rather small dimensions of the target. The values of the converter thickness for carrying out experimental studies at the linear electron accelerator "LINAC-300" of the National Scientific Center "Kharkov Institute of Physics and Technology" were obtained as a result of a computational experiment. It is shown that the flux of gamma quanta in front of the target is maximum at 2 mm of tungsten, but there are also high-energy electrons in front of the target. The electron flux in front of the target is much less for a 3 mm thick tungsten converter. The gamma quanta flux decreased slightly in this case.

The method of radiation damage preliminary estimation in the target for different converter thickness values, and, therefore, different values of the gamma quanta flux crossing the target boundaries, was investigated by the computer simulation method using Geant4 toolkit.

The possibility to optimize the experimental stand for studying the main mechanisms leading to the ruptures of organic dye molecules appeared on the basis of the obtained results. It is planned to receive specific proposals for optimizing the experimental stand in further research.

## ACKNOWLEDGEMENTS
The authors thank the Department of Astronomy & Astrophysics at the University of Toronto for the provided hardware calculating facilities, as well as Meest corporation and the "Computers for Ukraine" project for crucial logistical assistance in the transfer of these facilities to Ukraine.

**ORCID IDs:**
**Tetiana V. Malykhina** https://orcid.org/0000-0003-0035-2367; **Valentin I. Kasilov** https://orcid.org/0000-0002-1355-311X
**Vladimir E. Kovtun** https://orcid.org/0000-0001-8966-7685; **Sergey P. Gokov** https://orcid.org/0000-0002-3656-3804